\documentclass[12pt,preprint]{aastex}

\slugcomment{Astronomical Journal}
\shorttitle{NGC 4418}
\shortauthors{Imanishi et al.}

\begin{document}

\title{Near-infrared and Millimeter Constraints on the Nuclear Energy
Source of the Infrared Luminous Galaxy NGC 4418}

\author{Masatoshi Imanishi\altaffilmark{1}}
\affil{National Astronomical Observatory, 2-21-1, Osawa, Mitaka, Tokyo
181-8588, Japan} 
\email{imanishi@optik.mtk.nao.ac.jp} 

\author{Kouichiro Nakanishi, Nario Kuno}
\affil{Nobeyama Radio Observatory, Minamimaki, Minamisaku, Nagano,
384-1305, Japan} 
\email{nakanisi@nro.nao.ac.jp, kuno@nro.nao.ac.jp}

\and        

\author{Kotaro Kohno}
\affil{Institute of Astronomy, University of Tokyo, 2-21-1, Osawa, Mitaka, 
Tokyo, 181-0015, Japan}
\email{kkohno@ioa.s.u-tokyo.ac.jp}

\altaffiltext{1}{Visiting Astronomer at the Infrared Telescope Facility,
which is operated by the University of Hawaii under
Cooperative Agreement no. NCC 5-538 with the National
Aeronautics and Space Administration, Office of Space
Science, Planetary Astronomy Program.}

\begin{abstract}
We present near-infrared and millimeter investigations of the nucleus
of the infrared luminous galaxy NGC 4418, which previous observations
suggest possesses a powerful buried AGN. We found the following main
results: (1) The infrared $K$-band spectrum shows CO absorption
features at 2.3--2.4 $\mu$m owing to stars and very strong H$_{2}$
emission lines. The luminosity ratios of H$_{2}$ emission lines are
suggestive of a thermal origin, and the equivalent width of the H$_{2}$
1--0 S(1) line is the second largest observed to date in an external
galaxy, after the well-studied strong H$_{2}$-emitting galaxy NGC 6240.
(2) The infrared $L$-band spectrum shows a clear polycyclic aromatic
hydrocarbon (PAH) emission feature at 3.3 $\mu$m, which is usually
found in star-forming galaxies. The estimated star-formation
luminosity from the observed PAH emission can account for only a small
fraction of the infrared luminosity. (3) Millimeter interferometric
observations of the nucleus reveal a high HCN (1--0) to HCO+ (1--0)
luminosity ratio of $\sim$1.8, as has been previously found in pure
AGNs. (4) The measurements of HCN (1--0) luminosity using a
single-dish millimeter telescope show that the HCN (1--0) to infrared
luminosity ratio is slightly larger than the average, but within the
scattered range, for other infrared luminous galaxies. 
All of these results can be explained by the scenario in which, in
addition to energetically-insignificant, weakly-obscured star-formation 
at the surface of the nucleus, a powerful X-ray emitting AGN deeply
buried in dust and high density molecular gas is present.   

\end{abstract}


\section{Introduction}

Infrared luminous galaxies (L$_{\rm IR}$ $\gtrsim$
10$^{11}$L$_{\odot}$), discovered with the {\it IRAS} all sky
survey, radiate most of their luminosity as dust emission in the
infrared \citep{sam96}. Thus, powerful energy sources, either 
starbursts and/or active galactic nuclei (AGNs), are present 
hidden behind dust. As the bright end of the local luminosity function is
dominated by these infrared luminous galaxies \citep{soi87}, 
they constitute a very important population in the local universe. At
high redshift, these infrared luminous galaxies dominate the cosmic
infrared background emission, and have been used to trace the
dust-obscured star formation rate, dust content, and metallicity in
the early universe, based on the assumption that they are powered by
starbursts \citep{bar99}. Estimating the fraction of their infrared
luminosities that is powered by individual activity (i.e., starbursts
and AGNs) is fundamental not only to unveil the true nature of these
galaxies but also to understand obscured AGN-starburst connections in
the universe. As distant sources are too faint to investigate in
detail with existing observing facilities, a complete understanding of
these infrared luminous galaxies in the nearby universe is a first
step to achieving these goals.

If AGNs are present and obscured by dust in a {\it torus}
geometry, then clouds along the torus axis are photo-ionized by the
AGN's hard radiation (these are the so-called narrow line regions:
NLRs; Robson 1996). The NLRs produce optical emission lines with flux
ratios that are distinguishable from those of normal star-forming galaxies, so 
AGNs obscured by torus-shaped dust are detectable through optical
spectroscopy (classified as Seyferts; Veilleux, Kim, \& Sanders
1999a). Such obscured AGNs are also recognizable by detecting strong
high-excitation forbidden emission lines, originating in the NLRs, in
high-resolution near- to mid-infrared spectra \citep{gen98,vei99b}.
Approximately 10--20\% of nearby ($z <$ 0.3) infrared luminous
galaxies with L$_{\rm IR}$ $>$ 10$^{11}$L$_{\odot}$ are known to
possess such AGNs (Veilleux et al. 1999a,b). However, since the
nuclear regions of infrared luminous galaxies are very dusty
\citep{sam96}, AGNs resident in the majority of these galaxies may be
deeply embedded in dust along all sightlines. In such {\it buried}
AGNs, X-ray dissociation regions dominated by low-ionization species
\citep{mal96} develop, rather than NLRs, so that the signatures of
buried AGNs are difficult to find using the conventional methods
described above. A recent detailed investigation of the local infrared
60 $\mu$m luminosity function has indirectly suggested that buried
AGNs may be energetically important in infrared luminous galaxies
\citep{tak03}. It is desirable to establish a method by which to
reveal the presence of buried AGNs more directly and persuasively, and
to quantitatively determine their energetic importance in infrared
luminous galaxies.

Observations at wavelengths where dust extinction is low are clearly a
powerful way to investigate buried AGNs. Some methods have been
proposed, including thermal infrared (3--25 $\mu$m) imaging
\citep{soi00} and spectroscopic observations \citep{gen98,imd00},
millimeter interferometric observations \citep{koh02}, and high energy
X-ray observations \citep{fra03}. Some of these methods have
succeeded in providing strong evidence for buried AGNs in some
optically-non-Seyfert infrared luminous galaxies. Examples include Arp
299 \citep{del02}, NGC 4945 \citep{iwa93,don96}, IRAS 08572+3915
\citep{dud97,soi00,imd00}, and UGC 5101 \citep{idm01,im03,ima03}.
However, even in these galaxies, the properties of the energy sources
investigated using independent methods are not completely quantitatively
consistent, primarily because these sources may be AGN/starburst
composites, rather than pure buried AGNs or pure starbursts, and
because correction for dust extinction can cause some ambiguity. It is
important to observe selected buried AGN candidates in more detail, by
combining these potentially powerful methods, and to draw a
quantitatively consistent picture of their energy sources.

NGC 4418, at $z$ = 0.007, is one of the closest and strongest buried
AGN candidates. It has an infrared luminosity of L$_{\rm IR}$ $\sim$ 9
$\times$ 10$^{10}$L$_{\odot}$ (= 10$^{44.5}$ ergs s$^{-1}$; see Table
1), and shows no clear Seyfert signatures in its optical spectrum
\citep{arm89,leh95}. The 5--23 $\mu$m mid-infrared spectrum of NGC
4418 has the typical shape of an obscured AGN \citep{roc91,spo01}.
Based on an 8--23 $\mu$m mid-infrared spectrum, \citet{dud97} argued
that the energy source of NGC 4418 is very compact and more
centrally-concentrated than surrounding dust, as is expected for a
buried AGN. The core of NGC 4418 is spatially very compact
($\lesssim$0$\farcs$5) in the infrared and radio
\citep{eva03,con90,eal90,kew00}, again suggesting the presence of a
compact energy source like an AGN. From far-infrared spectroscopy, NGC
4418 is known to be an unusually weak [CII] (158 $\mu$m) emitter
relative to its far-infrared luminosity \citep{mal97,mal00}. The
presence of a powerful buried AGN is one explanation for this
\citep{mal97}, although some alternative scenarios exist
\citep{mal01}. In this way, evidence for a powerful buried AGN at the
center of NGC 4418 has been accumulated. However, the evidence still
may not be conclusive \citep{eva03}. Further constraints on the
nuclear energy source of NGC 4418 from the use of other techniques are
clearly warranted. In this paper, we report on near-infrared $K$-
(2.0--2.5 $\mu$m) and $L$-band (2.8--4.1 $\mu$m) spectroscopic
observations as well as millimeter (3.4 mm) observations using both a
single dish telescope and an interferometric array. Throughout this
paper, $H_{0}$ $=$ 75 km s$^{-1}$ Mpc$^{-1}$, $\Omega_{\rm M}$ = 0.3,
and $\Omega_{\rm \Lambda}$ = 0.7 are adopted, so that the distance to
NGC 4418 is 28.2 Mpc and 1 arcsec corresponds to 130 pc.

\section{Observations and Data Reduction}

\subsection{Near-infrared spectroscopy}

\subsubsection{Spectroscopy with SpeX at the IRTF 3-m telescope}

Near-infrared spectroscopic observations were performed with SpeX
\citep{ray03} at the IRTF 3-m telescope on Mauna Kea, Hawaii, on 2003
March 18 (UT). The 1.9--4.2 $\mu$m cross-dispersed mode with a
0\farcs8 wide slit was employed, so that both $K$- (2.0--2.5 $\mu$m)
and $L$-band (2.8--4.1 $\mu$m) spectra were taken simultaneously with
a resolution of R $\sim$ 1000. The sky conditions were photometric and
the seeing size at $K$ was measured to be $\sim$0$\farcs$6
(full-width at half-maximum; FWHM). The position angle of the slit was
set along the north-south direction. A standard telescope nodding
technique (ABBA pattern) was employed along the slit to subtract
background emission. We set the throw between the A and B positions to
be 7.5 arcsec, because infrared 2--20 $\mu$m emission from NGC 4418 is
dominated by the compact nucleus \citep{sco00,eva03}. With this throw,
the signal from the object was always inside the slit (15 arcsec long).

The telescope tracking was monitored with the infrared slit-viewer of
SpeX. Each exposure was 15 sec, and 2 coadds were employed at each
position. With this exposure time, signals at $\lambda_{\rm obs}$ $>$
4.1 $\mu$m (in the observed frame) exceed the linearity level of the
SpeX array, and so data at $\lambda_{\rm obs}$ $>$ 4.1 $\mu$m were
removed. One ABBA cycle results in a 2-min (15 sec $\times$ 2 coadds
$\times$ 4 positions) exposure time. This cycle was repeated 40 times
and so the total net on-source integration time was 80 min.

An F8-type main sequence star, HR 4708, was observed as a standard star,
with an airmass difference of $<$0.05 to NGC 4418, to correct for
the transmission of the Earth's atmosphere. The $K$- and $L$-band
magnitudes of HR 4708 were estimated to be $K$ = 5.04 and $L$ = 4.99
mag, respectively, from its $V$-band (0.6 $\mu$m) magnitude (6.40
mag), adopting the $V-K$ and $V-L$ colors of an F8 type main sequence
star \citep{tok00}.

Standard data reduction procedures were employed using IRAF
\footnote{IRAF is distributed by the National Optical
Astronomy Observatories, which are operated by the Association of
Universities for Research in Astronomy, Inc. (AURA), under cooperative
agreement with the National Science Foundation.}. 
First, bad pixels and pixels hit by cosmic rays were replaced with the
interpolated values of the surrounding pixels. Then, frames taken with
an A (or B) beam were subtracted from frames subsequently taken with a
B (or A) beam, and the resulting subtracted frames were added and then
divided by a spectroscopic flat image. The spectra of NGC 4418 and HR
4708 were then extracted by integrating signals over 2$\farcs$4 along
the slit. Wavelength calibration was performed using the
wavelength-dependent transmission of the Earth's atmosphere. The NGC
4418 spectrum was divided by that of HR 4708, and was multiplied by
the spectrum of a blackbody with T$_{\rm eff}$ = 6000 K, corresponding
to an F8-type main sequence star \citep{tok00}. Flux-calibration was
made based on the signals detected inside our slit and on the adopted
magnitudes of HR 4708.

In the $L$-band spectrum, S/N ratios in the continuum are much worse
than those in the $K$-band, because of the higher background signals
from the Earth's atmosphere in the former. Thus, appropriate spectral
binning to R $\sim$ 150 was applied. For an obscured AGN candidate
such as NGC 4418, the targeted features in the $L$-band are the 3.1
$\mu$m absorption feature caused by ice-covered dust grains, the 3.3
$\mu$m polycyclic aromatic hydrocarbon (PAH) emission feature, and the
3.4 $\mu$m absorption feature due to bare carbonaceous dust grains
\citep{im03}. Since all of these features are spectrally broad, the
adopted spectral resolution is sufficient to investigate their
properties.

\subsubsection{Spectroscopy with NIRC at the Keck I 10-m telescope}

Since NGC 4418 is a nearby source, the PAH emission, one of the
important features used in our scientific interpretation, has a peak
at 3.313 $\mu$m (= 3.29 $\mu$m $\times$ 1.007). At 3.31--3.32 $\mu$m a
deep methane absorption feature is present owing to the Earth's atmosphere, peaking
at 3.315 $\mu$m, so the transmission of signals from
celestial objects is extremely low and the background level is very
high, both of which make a spectrum very noisy around this wavelength
range. Thus, the determined PAH emission flux can be uncertain if
based only on a single spectrum. An independent spectrum taken at a
different time with a different instrument is important for a
quantitative discussion.

An $L$-band spectrum of NGC 4418 was also taken with NIRC
\citep{mat94} at the Keck I 10-m telescope on 2000 March 24 (UT), under
clear weather conditions. The LM + gr60 grism was used with a 4.5-pix
slit (0$\farcs$68 wide). The resulting spectral resolution is R $\sim$
70. The position angle of the slit was set to 35$^{\circ}$ east of
north. Signals were taken at five positions separated by 4$\farcs$8 along
the slit. Each exposure was 0.41 sec, and 100 coadds were employed at
each position. The total net on-source integration time was 41 min
(0.41 sec $\times$ 100 coadds $\times$ 5 positions $\times$ 12
cycles). HD 106965 (L = 7.295) was observed as a standard star, 
with an airmass difference of $<$0.05 to NGC 4418, to correct for the
transmission of the Earth's atmosphere.

Data reduction procedures were carried out using IRAF, in a similar
way as those employed for the IRTF SpeX spectrum. First, bias and dark
frames were subtracted from each frame, and then the successive frame
was subtracted. After flat-field and bad-pixel corrections had been
applied, the frames were shifted, and then summed to extract a
spectrum. Wavelength calibration was carried out using the
wavelength-dependent transmission of the Earth's atmosphere. The NGC
4418 spectrum was divided by that of HD 106965, multiplied by the
spectrum of a blackbody with T$_{\rm eff}$ = 8500 K, flux-calibrated,
and then a final spectrum was obtained.

\subsection{Millimeter observations}

\subsubsection{Nobeyama 45-m telescope}

We used the 45-m telescope of the Nobeyama Radio Observatory (NRO) to
measure the luminosity of HCN (1--0) ($\lambda_{\rm rest}$ = 3.3848 mm
or $\nu_{\rm rest}$ = 88.632 GHz). The half-power beam width (HPBW) of
the telescope was $\sim$19 arcsec, and the main beam efficiency was
$\sim$0.51 at this wavelength (frequency). The HCN observations were
made on 2002 November 28 (UT), under clear weather conditions. The
HCN-line data were taken simultaneously with two SIS receivers, S80
and S100, which can observe two orthogonal polarizations
simultaneously. However, the performance of S80 was unstable during
our observations, and thus we used only S100 data for our analysis.
The receiver backends were 2048 channel wide-band acousto-optical
spectrometers (AOSs). The frequency resolution and the total band
width were 250 kHz and 250 MHz, respectively, which correspond to 0.85
km s$^{-1}$ and 850 km s$^{-1}$ at the observed wavelength
(frequency). An intensity calibration was performed using the chopper
wheel method. The system noise temperatures (SSB), including the
antenna ohmic loss and the Earth's atmospheric effect, were 230--280 K
at the observed wavelength (frequency). The telescope pointing was
checked every 50 minutes by observing the bright quasar 3C 273. We
used a position-switching mode with an integration time of 20 sec for
on- and off-source.

The pointing of the NRO 45-m telescope is very sensitive to wind. To
obtain a reliable flux estimate, we used only data taken during calm
wind conditions (wind speed $\lesssim$ 10 m s$^{-1}$) for the
analysis. The total net on-source exposure time for the HCN
observations was 5600 sec.

Data analysis was made using the software Newstar and AIPS in a
standard manner. In each frame, the stability of the baseline was
checked by eye, and data with unstable baselines were removed. The
baseline was fit with a first order polynomial function and subtracted
from the actual spectrum to show the HCN line spectrum.

\subsubsection{RAINBOW Interferometer and Nobeyama Millimeter Array}

Interferometric observations of HCN (1--0) and HCO+ (1--0)
($\lambda_{\rm rest}$ = 3.3637 mm or $\nu_{\rm rest}$ = 89.188 GHz)
were performed with the RAINBOW Interferometer at NRO on 2004 January
27 and 28 (UT) under clear and virtually windless (wind speed $<$ 5 m
s$^{-1}$) observing conditions. The RAINBOW Interferometer is the
seven-element combined array that includes six 10-m antennas (Nobeyama
Millimeter Array; NMA) and the NRO 45-m telescope. 
During the daytime of the first observation (January 27), there was
snow fall, and frozen ice on the surface of the primary mirror of the
45-m telescope had not melted completely by the time of our night
observations. Thus, only six 10-m antennas (i.e., the NMA), with the AB
array configuration, were used for the observations. In this AB
configuration, the longest baseline was 350 m. On the second night,
all seven antennas (i.e., RAINBOW) were used, with a longest baseline
of 410 m.

The backend was the Ultra-Wide-Band Correlator, UWBC \citep{oku00}, which
was configured to cover 1024 MHz with 128 channels at 8 MHz
resolution. The central frequency was set to be 88.29 GHz, so that
both the HCN and HCO+ lines redshifted to 88.016 GHz and 88.569 GHz,
respectively, were simultaneously observable. The band width of 1024
MHz corresponds to $\sim$3500 km s$^{-1}$ at $\nu$ $\sim$ 88 GHz.
Since the Hanning window function was applied to reduce side lobes in
spectra, the actual resolution was widened to 16 MHz or 55 km s$^{-1}$
at $\nu$ $\sim$ 88 GHz. The bright quasar 3C 273 was used to calibrate
temporal variations in the visibility amplitude and phase, as well
as passband across the 128 channels. As the NRO 45-m telescope is very
sensitive to wind, the maser source R-Com was prepared to use as a
pointing check. However, wind was very weak on the second night and
the visibility amplitudes between the 10-m antennas and the 45-m antenna
were sufficiently large throughout the observations. No pointing check
was made for the 45-m telescope.

Standard data reduction was employed using the package UVPROC-II
developed at NRO \citep{tsu97}. For the second night's data, the
position dependence of the secondary mirror of the 45-m telescope on
elevation was first corrected. For both days' data, antenna baselines,
bandpass properties, and the time variation of visibility amplitude
and phase were corrected for, using the 3C~273 data. Data taken during
some fraction of the observing time showed large phase scatters, owing to
bad radio seeing. As these datasets are useless, they were removed.
After clipping a small fraction of unusually high amplitude data, the
data were Fourier-transformed using a natural {\it uv} weighting. The
flux-calibration of NGC 4418 was made by observing the quasar
1741$-$038, whose flux level relative to Uranus had been measured.
Since the declination of NGC 4418 is almost 0$^{\circ}$, strong side
lobes were evident along the north-south direction. A conventional
CLEAN method was applied to deconvolve the synthesized beam pattern.

\section{Results}

\subsection{Near-infrared Spectra}

\subsubsection{$K$-band}

Figure 1 shows a flux-calibrated infrared $K$-band slit spectrum of
NGC 4418, taken with IRTF SpeX. The signals integrated over 0$\farcs$8
$\times$ 2$\farcs$4 give $K$ = 12.6 mag, which is in reasonable
agreement with previously obtained photometric measurements using
small apertures; $K$ = 12.8 mag (1$\farcs$7) by \citet{eal90}, or $K$
= 13.3 mag (1$\farcs$1) and $K$ = 11.8 mag (5$''$) ({\it HST} NICMOS
measurements) by \citet{eva03}.
Thus, slit loss in our spectrum is expected to be insignificant.

Regarding the $K$-band spectrum of NGC 4418, only a low-resolution (R
$\sim$ 100) spectrum has previously been reported by \citet{rid94},
who argued that the $K$-band spectrum is typical of a normal
star-forming galaxy, with the exception of a strong H$_{2}$ 1--0 S(1)
emission line at $\lambda_{\rm rest}$ = 2.122 $\mu$m. Our higher
resolution (R $\sim$ 1000) spectrum newly reveals multiple H$_{2}$
emission lines, and enables quantitative measurements. Important
emission lines are indicated in Figure 1, and their properties are
summarized in Table 2. The rest-frame equivalent width of the H$_{2}$
1--0 S(1) emission was measured to be EW$_{\rm S(1)}$ = 36 $\rm \AA$,
which is significantly larger than in the majority of infrared
luminous galaxies ($\lesssim$20 $\rm \AA$; Goldader et al. 1997b), and
the second highest among galaxies so far known, next to NGC 6240 (=69
$\rm \AA$; Goldader et al. 1997b), the well-studied, unusually strong
H$_{2}$-emitting galaxy \citep{rie85}. The H$_{2}$ 1--0 S(1) to
infrared luminosity ratio is L$_{\rm S(1)}$/L$_{\rm IR}$ $\sim$
10$^{-5.4}$.  Despite the large equivalent width (EW$_{\rm S(1)}$),
the L$_{\rm S(1)}$/L$_{\rm IR}$ ratio is in the lowest scattered range
observed in infrared luminous galaxies (=10$^{-5.11 \pm 0.28}$;
Goldader et al.  1997a).

There is a gap in the continuum level of the spectrum between $\lambda
<$ 2.3 $\mu$m and $>$ 2.3 $\mu$m. We attribute the gap to CO
absorption features at $\lambda_{\rm rest}$ = 2.31--2.4 $\mu$m, which
are usually found in star-forming galaxies \citep{gol97b}. To estimate
the absorption strength, we basically follow the method proposed by
Doyon, Joseph, \& Wright (1994), who defined a spectroscopic CO index
as
\begin{eqnarray} {\rm CO_{spec}} & \equiv & -2.5log<R_{2.36}>,
\end{eqnarray} where $<$R$_{2.36}$$>$ is the average of actual signals
at $\lambda_{\rm rest}$ = 2.31--2.40 $\mu$m divided by a power-law
continuum (F$_{\rm \lambda}$ = $\alpha \times \lambda^{\beta}$)
extrapolated from shorter wavelengths.  A continuum level is
determined using data points at $\lambda_{\rm rest}$ = 2.0--2.29
$\mu$m, excluding obvious emission lines. When CO absorption is
present, $<$R$_{2.36}$$>$ should be less than unity and thus CO$_{\rm
spec}$ should have a positive value. The adopted continuum level is
shown as a solid line in Figure 1, and we obtain CO$_{\rm spec}$ =
0.29. This value is as large as those of many star-forming galaxies
\citep{coz01} and slightly larger than the previous estimate for this
galaxy based on a lower resolution (R $\sim$ 100) spectrum
(=0.23$\pm$0.03; Ridgway et al. 1994). 

\subsubsection{$L$-band}

Figures 2a and 2b show flux-calibrated $L$-band spectra taken with
IRTF SpeX and Keck NIRC, respectively. The 3.3 $\mu$m PAH emission
feature is found in both spectra. Its line width appears to be larger
in the NIRC spectrum owing to its lower spectral resolution. The SpeX
and NIRC spectra give $L$ = 12.1 mag (0$\farcs$8 $\times$
2$\farcs$4) and $L$=12.3 (0$\farcs$7 $\times$ 4$''$), respectively.

From our slit spectrum taken with SpeX, we find the nuclear $K-L$
color to be 0.5 mag. In the case of SpeX data, both $K$- and $L$- band
spectra were taken simultaneously under the same observing conditions,
toward the same regions of the galaxy, with a similar level of
possible slit loss, so that the measured $K-L$ color of 0.5 mag can be
taken to be more certain than the absolute $K$- and $L$-band
magnitudes.  The nuclear $K-L$ color is typical of star-forming
galaxies \citep{wil84}.

The flux of the 3.3 $\mu$m PAH emission was estimated using the
spectral profile of type-1 sources \citep{tok91} as a template for the
3.3 $\mu$m PAH emission, as was made by \citet{ima02,ima03a}. The SpeX
and NIRC spectra give a 3.3 $\mu$m PAH emission flux of f$_{\rm
3.3PAH}$ $\sim$ 6 and 8 $\times$ 10$^{-14}$ ergs s$^{-1}$ cm$^{-2}$,
respectively. We adopt f$_{\rm 3.3PAH}$ = 7 $\times$ 10$^{-14}$ ergs
s$^{-1}$ cm$^{-2}$, in which case the luminosity and rest-frame
equivalent width of the 3.3 $\mu$m PAH emission are L$_{\rm 3.3PAH}$
$\sim$ 7 $\times$ 10$^{39}$ ergs s$^{-1}$ and EW$_{\rm 3.3PAH}$ $\sim$
70 nm, respectively. The spectral shape of NGC 4418, with a large
EW$_{\rm 3.3PAH}$ ($\sim$ 70 nm), is also typical of star-forming
galaxies \citep{imd00}.

\subsection{Millimeter Data}

\subsubsection{Single dish data}

Figure 3 shows a spectrum of the HCN (1--0) line taken with the NRO
45-m telescope. Using a Gaussian fit, we estimate the HCN flux to be
3.5 K km s$^{-1}$. The HCN flux, estimated based on a simple summation
of signals above an adopted continuum level, gives a value consistent
within 15\%. By adopting the Jy-to-K conversion factor of the NRO 45-m
telescope ($\sim$ 2.4 Jy K$^{-1}$), we obtain an HCN flux of 8.5 Jy km
s$^{-1}$. Using equation (1) of \citet{gao04a}, we estimate the HCN
luminosity to be 2.6 $\times$ 10$^{7}$ K km s$^{-1}$ pc$^{2}$. Figure
4 compares the infrared to HCN luminosity ratio of NGC 4418 with those
of other infrared luminous galaxies. The ratio for NGC 4418 is
slightly higher than the average, but within the scattered range, of
the ratios found in infrared luminous galaxies.

\subsubsection{Interferometric data}

A nuclear spectrum before continuum subtraction taken with the
interferometric array showed that the flux level in between the HCN (1--0)
and HCO+ (1--0) emission lines is significantly above the zero level,
suggesting that continuum emission is detected. We combined data
points that were unaffected by the HCN and HCO+ emission lines and
estimated the continuum flux to be $\sim$8.5 mJy at $\lambda$ $\sim$
3.4 mm ($\nu$ $\sim$ 88 GHz). Figure 5 displays the contours of the
continuum emission. The continuum emission is spatially compact, and
the peak position is spatially coincident with the optical nucleus
(J1950; 12$^{h}$26$^{m}$54.62$^{s}$,
$-$00$^{\circ}$52$'$39$\farcs$20).

After the estimated continuum had been subtracted, we investigated the HCN
and HCO+ emission lines from NGC 4418. Figure 6 shows the channel map
around the HCN and HCO+ lines. The HCN and HCO+ emission lines are
also spatially compact and are spatially coincident with the optical
peak.

Figure 7 shows a continuum-subtracted nuclear spectrum around the HCN
and HCO+ emission lines. The integrated fluxes of HCN and HCO+ at the
peak position are estimated to be S$_{\rm HCN}$ = 10.3 Jy km s$^{-1}$
and S$_{\rm HCO+}$ = 5.6 Jy km s$^{-1}$, respectively. We thus obtain
an HCN-to-HCO+ flux ratio of 1.84 at the NGC 4418 nucleus. 
Although the {\it absolute} flux calibration uncertainty of the RAINBOW
data can be as large as $\sim$20\%, the HCN-to-HCO+flux {\it ratio}
should be very reliable, because both lines were observed simultaneously
with the same receiver and same correlator unit. 

The estimated HCN flux values based on the interferometric array (=10.3 Jy km
s$^{-1}$) and on the NRO 45-m telescope (=8.5 Jy km s$^{-1}$) agree
within the absolute flux calibration uncertainty ($\sim$20\%),
suggesting that the bulk of the HCN emission from NGC 4418 is
spatially compact and is detected with the interferometric data.

\section{Discussion}

\subsection{Nuclear star formation}

The CO absorption features found in the near-infrared $K$-band
spectrum originate in stars and not an AGN \citep{iva00}. The 3.3
$\mu$m PAH emission feature also comes from star-forming activity,
rather than a pure AGN \citep{moo86,imd00}. The nuclear $K - L$ color
of 0.5 mag measured with SpeX data is similar to that of star-forming galaxies
\citep{wil84}, and much bluer than the typical colors observed in many
Seyfert nuclei \citep{alo03,ia04}. All of these results suggest that
star-forming activity is the dominant contributor to the observed
nuclear $K$- and $L$-band fluxes, as was argued by \citet{spo01} based
on the infrared spectral energy distribution of NGC 4418.

The nature of the detected nuclear star formation in NGC 4418 can be
investigated using the observed 3.3 $\mu$m PAH emission luminosity. In
NGC 4418, the 3.3 $\mu$m PAH luminosity (= 7 $\times$ 10$^{39}$ ergs
s$^{-1}$) comes from regions at $\sim$2 arcsec$^{2}$ or $\sim$0.033
kpc$^{2}$. Thus, its surface brightness is S$_{\rm 3.3 PAH}$ $\sim$ 2
$\times $ 10$^{41}$ ergs s$^{-1}$ kpc$^{-2}$.
This corresponds to the infrared surface brightness with $\sim$2
$\times$ 10$^{44}$ ergs s$^{-1}$ kpc$^{-2}$ \citep{mou90,ima02}, or 
the estimated star formation rate per unit area with 
$\sim$9 M$_{\odot}$ yr$^{-1}$ kpc$^{-2}$ \citep{ken98}.
This value substantially exceeds the lower
limit of 10$^{-1}$ M$_{\odot}$ yr$^{-1}$ kpc$^{-2}$ beyond which
superwind activity, a characteristic property of starbursts, can occur
\citep{hec01}. We thus conclude that the detected 3.3 $\mu$m PAH
emission comes from nuclear starbursts, rather than central regions of
quiescent star-forming activity in the host galaxy.

The measured 3.3 $\mu$m PAH to infrared luminosity ratio in NGC 4418 is
$\sim$2 $\times$ 10$^{-5}$, which is a factor of $\sim$50 smaller than
starburst-dominated galaxies ($\sim$10$^{-3}$; Mouri et al. 1990;
Imanishi 2002), suggesting that the detected nuclear starbursts, with
modest dust extinction in the near-infrared 3--4 $\mu$m, contribute
very little to the total infrared luminosity of NGC 4418. Since the
infrared luminosity is likely to come from the compact nuclear core
\citep{eva03}, rather than extended star formation in the host galaxy,
the dominant energy source, an AGN and/or very compact starbursts,
should be deeply buried at the core.

\subsection{The Nature of the Buried Energy Source: Signatures of an
AGN}

Figure 8 shows a plot of the HCN (1--0) to CO (1--0) ($\lambda_{\rm
rest}$ = 2.6 mm) and HCN (1--0) to HCO+ (1--0) luminosity ratios of
NGC 4418 and other nearby galaxies. Empirically, a pure AGN (with no
detectable nuclear star-forming activity) shows high HCN/HCO+ and
HCN/CO luminosity ratios, as compared to star-forming galaxies
\citep{koh02,koh03}. Since HCN traces much higher-density molecular gas
(n$_{\rm H}$ $\sim$ 10$^{4}$ cm$^{-3}$) than is traced by CO (n$_{\rm
H}$ $\sim$ 10$^{2}$ cm$^{-3}$), the HCN/CO ratio can vary depending on
the density of molecular gas. However, both HCN and HCO+ probe
molecular gas of similar density (n$_{\rm H}$ $\sim$ 10$^{4}$
cm$^{-3}$), so that the HCN/HCO+ ratio is robust to the density of
molecular gas. The high HCN/HCO+ ratio in an AGN may be owing to the
enhancement of HCN in molecular gas irradiated by strong X-ray
emission from the AGN \citep{lep96}. Although a full explanation for
this high HCN/HCO+ ratio in an AGN is still to be developed, this
empirical method can be used to find a buried AGN, because of
negligible dust extinction in the millimeter wavelength range. The
high HCN/HCO+ luminosity ratio at the nucleus of NGC 4418 suggests the
presence of a buried AGN.

The strong CO absorption in the near-infrared $K$-band, and strong 3.3
$\mu$m PAH emission in the $L$-band, suggest that emission from normal
star-formation has an important contribution to the observed fluxes in
these wavelength ranges. However, the H$_{2}$ emission in the $K$-band
spectrum is unusually strong as compared to that of normal star-forming galaxies
\citep{rid94}. In normal star-forming galaxies, the H$_{2}$ 1--0 S(1)
to Br$\gamma$ emission-line luminosity ratios (L$_{\rm S(1)}$/L$_{\rm
Br\gamma}$) are smaller than or comparable to unity
\citep{mou89,mou92,gol97b,coz01}. The L$_{\rm S(1)}$/L$_{\rm
Br\gamma}$ ratio of $>$8 (Table 2) in NGC 4418 is unusually large.
Together with the large H$_{2}$ equivalent width (EW$_{\rm S(1)}$) of
NGC 4418, this suggests that in addition to emission from normal
star-forming activity, some mechanism that can produce strong H$_{2}$
emission should exist at the nucleus of NGC 4418.

Figure 9 shows the luminosity ratios of some H$_{2}$ emission lines.
The small observed 2--1 S(1) to 1--0 S(1) luminosity ratio with $<$0.3
is incompatible with the non-thermal (UV fluorescence) model as the
primary emission mechanism for the detected strong H$_{2}$ emission,
and favors H$_{2}$ emission of thermal origin. Based on the available
H$_{2}$ data alone, some thermal models, including shocks, X-ray
heating, and UV, are not clearly distinguishable. However, the
measured HCN/HCO+ ratio can help to distinguish among these scenarios.

In the case of the strong H$_{2}$-emitting galaxy NGC 6240, the
H$_{2}$ emission was found to be of thermal origin \citep{sug97}, and
to come from regions between the double nucleus of NGC 6240
\citep{sug97}. Shocks driven by a galaxy-merger or superwind have been
proposed as the primary mechanism of the H$_{2}$ emission
\citep{rie85,van93,ohy00,ohy03}. From these shock regions, a low
HCN/HCO+ ratio with $\sim$0.67 was observed \citep{nak04}. The low
HCN/HCO+ ratio in shock regions may be due to the enhancement of HCO+
\citep{raw00,raw04}. However, NGC 4418 is a single nucleus source,
with no clear signs of recent galaxy-merger/interaction
\citep{eal90,eva03}. Furthermore, at the nucleus of NGC 4418, a high
HCN/HCO+ ratio was measured, as found in pure AGNs. Given that stellar
UV heating of molecular gas is also expected to show a low HCN/HCO+
ratio \citep{koh02}, X-ray heating of H$_{2}$ molecules caused by the
putative buried AGN \citep{dra90} seems plausible. The observed
H$_{2}$ line luminosity ratios are consistent with this X-ray heating
scenario, if the H$_{2}$ emission suffers from dust extinction with
A$_{\rm K}$ $\sim$ several mag (see Fig.9).

NGC 4418 is known to show a very strong 9.7 $\mu$m silicate dust
absorption feature, whose optical depth was measured to be
$\tau_{9.7}$ = 5.6--7 \citep{roc86,dud97,spo01}. Adopting the relation
of $\tau_{9.7}$/A$_{\rm V}$ = 0.05--0.1 established in the Galactic
interstellar medium \citep{roc84,roc85}, we obtain a dust extinction
toward the $\sim$10 $\mu$m continuum emitting regions of $A_{\rm V}$
$>$ 50 mag, or A$_{\rm K}$ $>$ 5 mag \citep{rie85a}. In the case of a
dust envelope around a buried AGN (= a centrally concentrated energy
source), dust has a strong temperature gradient in that inner (outer)
dust has a higher (lower) temperature. The 10 $\mu$m continuum
typically comes from 300-K dust, which is located in the outer parts,
compared to the innermost dust sublimation radius (T = 1000--1500
K). Thus, the estimated A$_{\rm V}$ traces the dust column toward some
outer part of the dust envelope around the buried AGN. H$_{2}$
emission excited by X-rays also comes from some outer regions of the dust
envelope, similarly to the 300-K dust. If we adopt a value for the dust
extinction toward the H$_{2}$-emitting regions of A$_{\rm K}$ $>$ 5
mag, the model prediction of X-ray heating in Figure 9 will go down,
and move slightly to the right, which is consistent with the current
observational constraints on the H$_{2}$ emission line luminosity
ratios.

The scenario of X-ray heating by the buried AGN can also explain the
following observational results: (1) the large H$_{2}$ S(1) to
Br$\gamma$ luminosity ratio, (2) the large equivalent width of the
H$_{2}$ 1--0 S(1) emission line (EW$_{\rm S(1)}$), and (3) the small
H$_{2}$ S(1) to infrared luminosity ratio (L$_{\rm S(1)}$/L$_{\rm
IR}$). On point (1), AGNs generally show significantly larger H$_{2}$
1--0 S(1) to Br$\gamma$ luminosity ratios (L$_{\rm S(1)}$/L$_{\rm Br
\gamma}$ $>$ 1) than do star-forming galaxies ($<$1; Mouri \& Taniguchi
1992). On point (2), the H$_{2}$ equivalent width (EW$_{\rm S(1)}$) of
NGC 4418 is in the highest range observed for AGNs \citep{mou92}. In a
buried AGN, the covering fraction of the central AGN by molecular gas
is higher than that of a typical AGN surrounded by torus-shaped
molecular gas. This higher covering fraction can produce stronger
X-ray heated H$_{2}$ emission. Additionally, the near-infrared
$K$-band continuum from an AGN originates mostly from 1000- to 1500-K
dust close to the innermost, dust sublimation radius \citep{bar87}.
The high obscuration toward the central AGN in NGC 4418, as is implied
from the presence of one of the strongest 9.7 $\mu$m silicate dust
absorption features observed \citep{roc91}, can significantly
attenuate the AGN-originated $K$-band continuum flux. The flux
attenuation of H$_{2}$ emission from some outer part of the dust
envelope is slightly smaller than the $K$-band continuum, producing a
higher H$_{2}$ equivalent width than in the case of unobscured AGNs. 
Finally, on point (3), since the H$_{2}$ emission from the buried AGN in
NGC 4418 is likely to be more highly dust obscured than typical AGNs,
the small L$_{\rm S(1)}$/L$_{\rm IR}$ ratio can be explained by dust
extinction. 

With dust extinction of A$_{\rm K}$ $>$ 5 mag, the dereddened H$_{2}$
1--0 S(1) luminosity becomes $>$1.3 $\times$ 10$^{41}$ ergs s$^{-1}$,
and the dereddened L$_{\rm S(1)}$ to L$_{\rm IR}$ ratio is $>$4
$\times$ 10$^{-4}$. The intrinsic X-ray luminosity that produces this
H$_{2}$ 1--0 S(1) luminosity is estimated to be L$_{\rm X}$(1--10keV)
$>$ 1 $\times$ 10$^{44}$ ergs s$^{-1}$ \citep{lep83}. The resulting
1--10 keV X-ray to infrared luminosity ratio is L$_{\rm
X}$(1--10keV)/L$_{\rm IR}$ $>$ 0.3, which is as high as those in
AGN-powered infrared luminous galaxies \citep{ris00}. Thus, in this
X-ray heated H$_{2}$ emission scenario, the putative buried AGN could
account for the bulk of the infrared luminosity from NGC 4418.

In summary, all the data newly available from our observations,
including the high HCN/HCO+ ratio, H$_{2}$ emission line luminosity
ratios, large H$_{2}$ to Br$\gamma$ luminosity ratio, and large H$_{2}$
equivalent width, can be consistently explained by the buried AGN
scenario.

The infrared to HCN luminosity ratio of NGC 4418 is located in the
highest part of the scattered range of infrared luminous starburst
galaxies (Figure 4). If this ratio is similar to that of starburst
galaxies, it is sometimes used to argue that the infrared luminosity
of a source is starburst-powered \citep{gao04b}. However, it simply
indicates that much of the infrared emission arises from dust in dense
molecular clouds (probed by HCN) whatever the energy source of the
luminosities may be \citep{bar97,im03}. The buried AGN in NGC 4418 is
most likely to be surrounded by high density gas, as is the case for
stars at the star-forming cores of starburst galaxies, and so the
similar infrared-to-HCN ratio of NGC 4418 to that of starburst
galaxies is quite reasonable.

The presence of buried AGNs in some non-Seyfert ultraluminous infrared
galaxies (ULIRGs) with L$_{\rm IR}$ $>$ 10$^{12}$L$_{\odot}$ has been
revealed from $L$-band spectra, based on significantly lower
equivalent widths of the 3.3 $\mu$m PAH emission feature (EW$_{\rm
3.3PAH}$) than starbursts, with strong absorption features
\citep{imd00,idm01,im03}, whereas NGC 4418 shows no such signature in
the $L$-band spectrum. This $L$-band spectroscopic method to find a
buried AGN is possible only if the AGN-powered dust emission 
contributes significantly to the observed flux. In ULIRGs, this is
often the case \citep{far03}. However, in the case of NGC 4418, the
observed $L$-band flux is unfortunately dominated by stellar emission,
rather than dust emission \citep{spo01}, making it difficult to
find the buried AGN signatures in the $L$-band spectrum.

The buried AGN scenario in NGC 4418 presumes the presence of an
intrinsically strong X-ray emitting energy source. No significant
X-ray emission was detected at E $<$ 10 keV \citep{mai03}. It has been
found that AGNs in roughly half of Seyfert-2 galaxies suffer from
Compton-thick (N$_{\rm H}$ $>$ 10$^{24}$ cm$^{-2}$) absorption
\citep{ris99}. In a Compton-thick AGN, the observed 2--10 keV emission is
dominated by a scattered/reflected component, rather than a direct,
transmitted component, making the observed 2--10 keV flux very weak.
The significantly deeper 9.7 $\mu$m silicate dust absorption feature
in NGC 4418 as compared to typical Seyfert-2 galaxies \citep{roc91} makes
it very likely that the putative AGN in NGC 4418 is obscured by
Compton-thick absorbing material. Sensitive X-ray observations at E
$>$ 10 keV are most desirable to unveil the presence of such a
Compton-thick buried AGN, as found in a few non-Seyfert infrared
luminous galaxies, NGC 4945 \citep{iwa93}, Arp 299 \citep{del02}, and
NGC 6240 \citep{vig99}.

\section{Summary}

We investigated the nuclear energy source of the infrared luminous
galaxy NGC 4418 through near-infrared $K$- and $L$-band spectroscopy,
and millimeter observations at $\lambda$ $\sim$ 3.4 mm using both a
single dish telescope and an interferometric array. Various previous
observations have suggested that this galaxy 
has a deeply buried AGN at the center. In the near-infrared
spectra, we detected clear PAH emission and CO absorption features,
suggesting that the observed near-infrared flux comes mostly from
star-forming activity. The magnitude of this detected star-forming
activity was measured from the observed PAH emission luminosity and
was found to account for only 1/50 of the infrared luminosity of this
galaxy. We therefore suggested that the primary energy source of NGC
4418 is deeply buried in dust.

In the near-infrared $K$-band spectrum and millimeter interferometric
data, large-equivalent-width H$_{2}$ emission and a high HCN/HCO+
luminosity ratio, respectively, were found, both of which are
naturally explained by the presence of a strong X-ray emitting source,
such as an AGN, at the center of NGC 4418. The measured HCN to
infrared luminosity ratio in NGC 4418 can also be explained by the AGN
scenario, if the AGN is deeply buried in high-density molecular gas
and dust. Overall, our new near-infrared and millimeter data supported 
the buried AGN scenario in NGC 4418.

\acknowledgments

We are grateful to S. B. Bus, P. Sears (IRTF), R. Campbell, C. Sorenson,
W. Wack (Keck), H. Mikoshiba, Y. Iizuka, S. Ishikawa, K. Miyazawa,
C. Miyazawa, and A. Sakamoto (NRO) for their support during our
observing runs.
NRO is a branch of the National Astronomical Observatory, National
Institutes of Natural Sciences, Japan.
Some parts of the data analysis were made using a computer system operated
by the Astronomical Data Analysis Center (ADAC) and the Subaru Telescope
of the National Astronomical Observatory.
Some of the data presented herein were obtained at the W.M. Keck
Observatory, while MI was staying at the Institute for Astronomy,
University of Hawaii.
MI is grateful to A. T. Tokunaga for giving him the opportunity for the stay,
which was financially supported by the Japan Society for the Promotion of
Science.
The W.M. Keck Observatory is operated as a scientific partnership among the
California Institute of Technology, the University of California and the
National Aeronautics and Space Administration. The observatory was made
possible by the generous financial support of the W.M. Keck Foundation.
The authors wish to recognize and acknowledge the very significant
cultural role and reverence that the summit of Mauna Kea has always had
within the indigenous Hawaiian community.  We are most fortunate to have
the opportunity to conduct observations from this mountain.

\clearpage


\begin{deluxetable}{ccccccc}
\tablecaption{Detailed information on NGC 4418 \label{tab1}}
\tablewidth{0pt}
\tablehead{
\colhead{Object} & \colhead{Redshift}   & 
\colhead{f$_{\rm 12}$}  & \colhead{f$_{\rm 25}$}  & 
\colhead{f$_{\rm 60}$}  & \colhead{f$_{\rm 100}$}  & 
\colhead{log L$_{\rm IR}$} \\
\colhead{} & \colhead{}   & \colhead{(Jy)} & \colhead{(Jy)}  & \colhead{(Jy)} 
& \colhead{(Jy)}  & \colhead{(ergs s$^{-1}$)}  \\
\colhead{(1)} & \colhead{(2)} & \colhead{(3)} & \colhead{(4)} & 
\colhead{(5)} & \colhead{(6)} & \colhead{(7)}
}
\startdata
NGC 4418 & 0.007 &  0.9 & 9.3 & 40.7 & 32.8 & 44.5 \\
\enddata

\tablecomments{
Column (1): Object.
Column (2): Redshift. 
Columns (3)--(6): f$_{12}$, f$_{25}$, f$_{60}$, and f$_{100}$ are 
{\it IRAS FSC}
fluxes at 12$\mu$m, 25$\mu$m, 60$\mu$m, and 100$\mu$m, respectively.
Column (7): Logarithm of infrared (8$-$1000 $\mu$m) luminosity
in ergs s$^{-1}$ calculated with
$L_{\rm IR} = 2.1 \times 10^{39} \times$ D(Mpc)$^{2}$
$\times$ (13.48 $\times$ $f_{12}$ + 5.16 $\times$ $f_{25}$ +
$2.58 \times f_{60} + f_{100}$) ergs s$^{-1}$
\citep{sam96}. 
}

\end{deluxetable}

\begin{deluxetable}{cccccc}
\tablecaption{Emission lines in the $K$-band \label{tab2}}
\tablewidth{0pt}
\tablehead{
\colhead{Line} & \colhead{$\lambda_{\rm rest}$}  & 
\colhead{Flux \tablenotemark{a}} & \colhead{Luminosity} 
& \colhead{FWHM}  & \colhead{Rest EW}  \\
\colhead{} & \colhead{($\mu$m)} & 
\colhead{($\times$10$^{-15}$ ergs s$^{-1}$ cm$^{-2}$)} 
& \colhead{($\times$10$^{38}$ ergs s$^{-1}$)} 
& \colhead{(km s$^{-1}$)} & \colhead{($\rm \AA$)} \\
\colhead{(1)} & \colhead{(2)} & \colhead{(3)} & \colhead{(4)} & 
\colhead{(5)} & \colhead{(6)}
}
\startdata
H$_{2}$ 1--0 S(2) & 2.034 &  3.4 & 3.2 & 220 & 8 \\
H$_{2}$ 1--0 S(1) & 2.122 & 13.2 &  13 & 350 & 36 \\
H$_{2}$ 1--0 S(0) & 2.223 &  3.9 & 3.7 & 260 & 12 \\
H$_{2}$ 1--0 Q(1) & 2.407 & 11.4 &  11 & 250 & 47 \\
H$_{2}$ 1--0 Q(3) & 2.424 &  9.5 & 9.0 & 300 & 39  \\
H$_{2}$ 2--1 S(3) & 2.074 & $<$3.8 & $<$3.6 & 500 \tablenotemark{b} & $<$11 \\
H$_{2}$ 2--1 S(1) & 2.248 & $<$3.2 & $<$3.0 & 500 \tablenotemark{b} & $<$10 \\
Br$\gamma$        & 2.166 & $<$1.7 & $<$1.6 & 500 \tablenotemark{b} & $<$6 \\
\enddata

\tablecomments{
Column (1): Line.
Column (2): Rest-frame wavelength. 
Column (3): Observed flux.
Column (4): Observed luminosity. 
Column (5): Full width at half maximum.
Column (6): Rest-frame equivalent width.
}

\tablenotetext{a}{The flux is estimated from a Gaussian fit. 
We also estimate the flux by simply summing signals above an adopted
continuum level, which gives values consistent within $\sim$20\% for all
the detected lines.}

\tablenotetext{b}{The line width is conservatively assumed to be 500 km
s$^{-1}$ to estimate the upper limit of the flux, luminosity, and
rest-frame equivalent width of an emission line.}

\end{deluxetable}

\begin{figure}
\includegraphics[angle=-90,scale=.5]{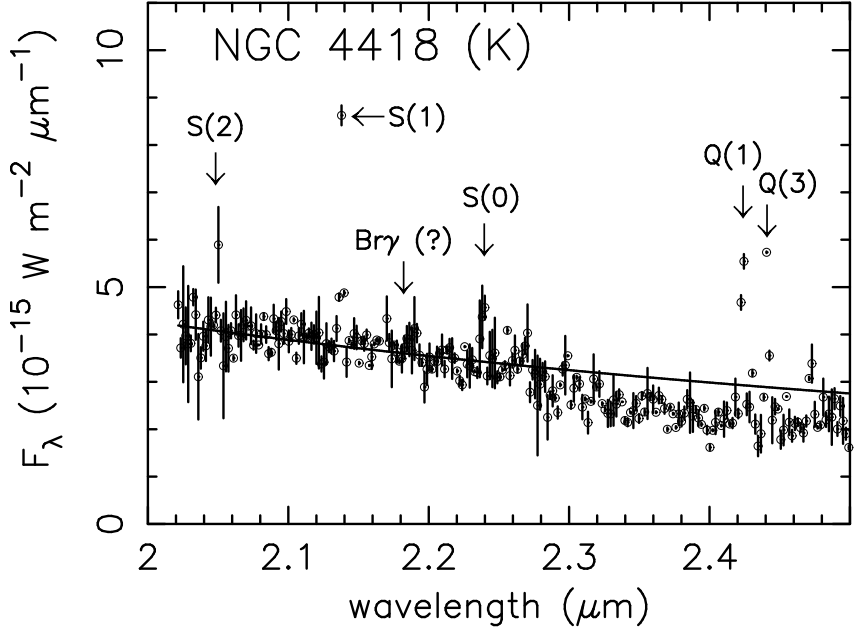}
\caption{
Near-infrared $K$-band slit spectrum of the NGC 4418 nucleus, taken with
IRTF SpeX. 
The abscissa and ordinate are the observed wavelength in $\mu$m and
flux F$_{\lambda}$ in 10$^{-15}$ W m$^{-2}$ $\mu$m$^{-1}$, respectively.
The solid line is the adopted continuum level with respect to which 
the strength of the CO absorption features is measured. 
Some detected H$_{2}$ emission lines are indicated.
S(2): H$_{2}$ 1--0 S(2) at $\lambda_{\rm rest}$ = 2.034 $\mu$m.
S(1): H$_{2}$ 1--0 S(1) at $\lambda_{\rm rest}$ = 2.122 $\mu$m.
S(0): H$_{2}$ 1--0 S(0) at $\lambda_{\rm rest}$ = 2.223 $\mu$m.
Q(1): H$_{2}$ 1--0 Q(1) at $\lambda_{\rm rest}$ = 2.407 $\mu$m.
Q(3): H$_{2}$ 1--0 Q(3) at $\lambda_{\rm rest}$ = 2.424 $\mu$m.
The Br$\gamma$ emission line at 
$\lambda_{\rm rest}$ = 2.166 $\mu$m is not clearly detected. 
Its expected wavelength is marked with ``Br$\gamma$ (?)''. 
}
\end{figure}

\begin{figure}
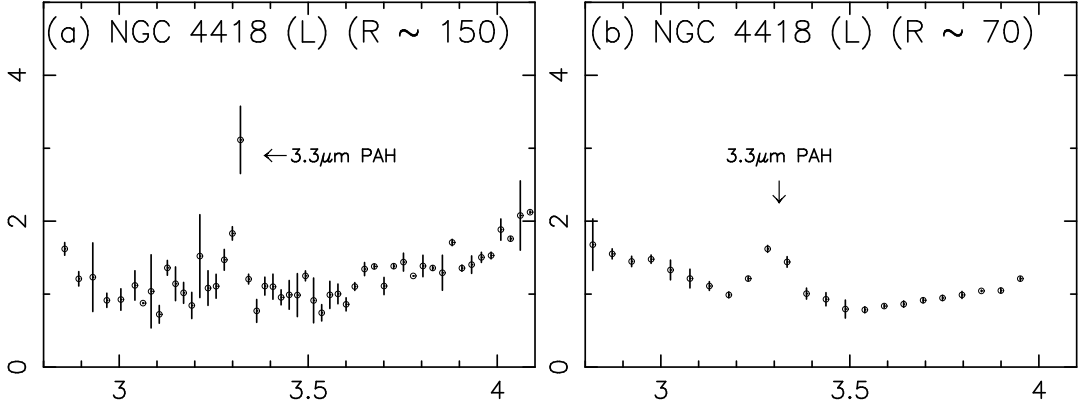

\includegraphics[angle=-90,scale=.35]{Imanishi.fig2a.eps}
\includegraphics[angle=-90,scale=.35]{Imanishi.fig2b.eps}
\caption{
Near-infrared $L$-band slit spectrum of the NGC 4418 nucleus.
The abscissa and ordinate are the observed wavelength in $\mu$m and
flux F$_{\lambda}$ in 10$^{-15}$ W m$^{-2}$ $\mu$m$^{-1}$, respectively.
{\it (a)}: Taken with IRTF SpeX with an aperture of 0$\farcs$8 $\times$
2$\farcs$4. 
The spectral resolution is binned to R $\sim$ 150 at $\lambda$ = 3.3
$\mu$m.  
The large error bar at the peak position of the 3.3 $\mu$m PAH emission
is likely to be caused by methane absorption at 3.315 $\mu$m by the
Earth's atmosphere (see \S 2.1.2). 
{\it (b)}: Taken with Keck I NIRC with an aperture of 0$\farcs$7 $\times$
4$''$. The spectral resolution is R $\sim$ 70.
}
\end{figure}

\begin{figure}
\includegraphics[angle=-90,scale=.5]{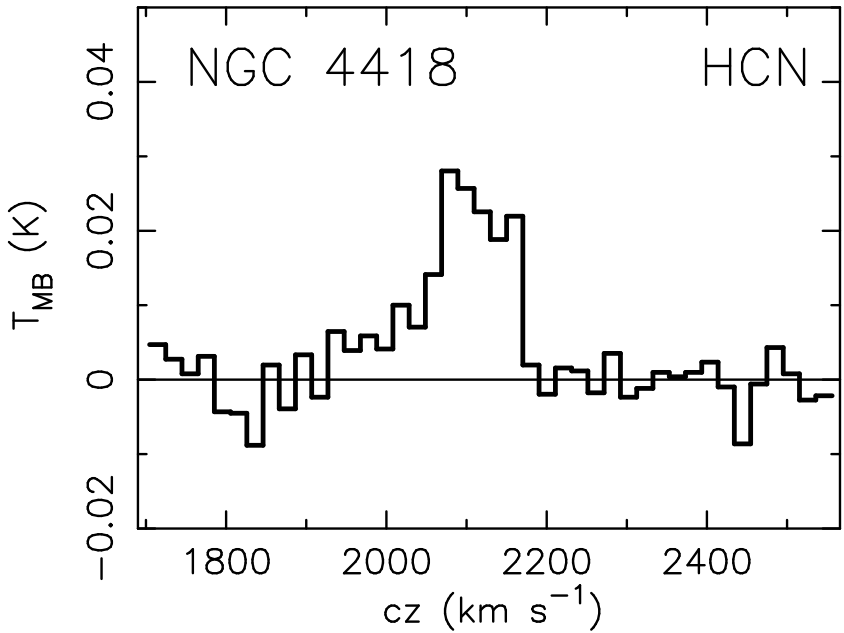}
\caption{
HCN (1--0) emission from NGC 4418 observed with the S100 receiver of the
NRO 45m telescope (19 arcsec half-power beam width). 
The abscissa is velocity in km s$^{-1}$, and the ordinate is main beam
temperature in K. 
The spectrum is smoothed to a velocity resolution of 20 km s$^{-1}$.
}
\end{figure}

\begin{figure}
\includegraphics[angle=-90,scale=.6]{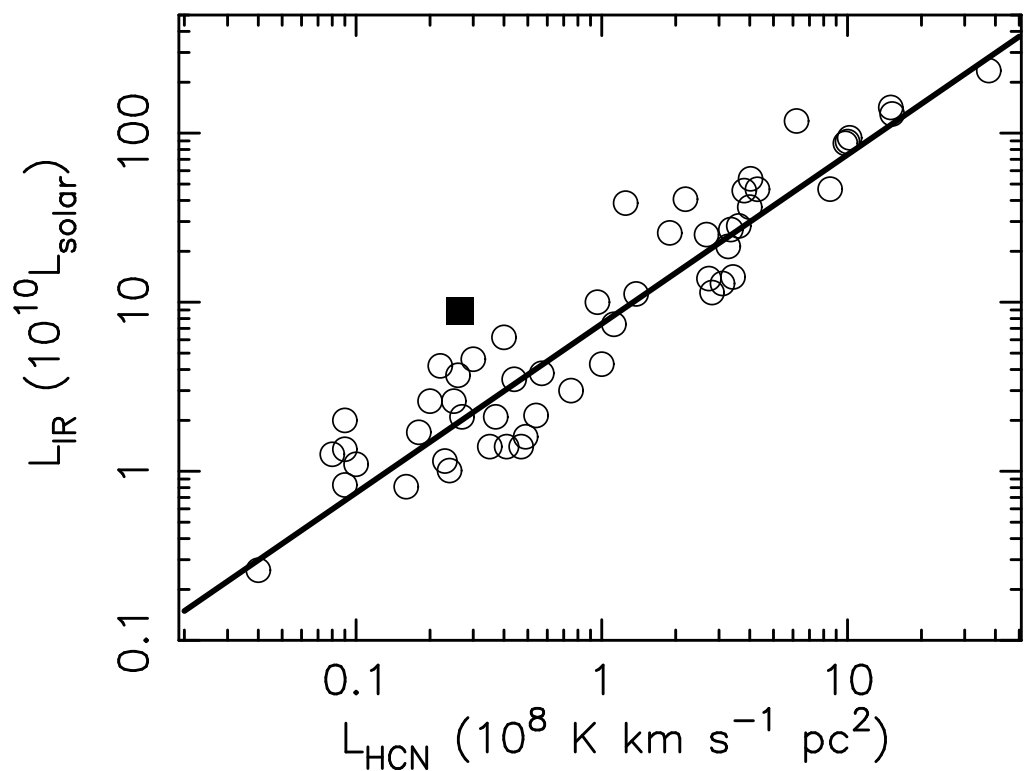}
\caption{
Relation between HCN (1--0) and infrared luminosity.
The abscissa is HCN luminosity in 10$^{8}$ K km s$^{-1}$ pc$^{-2}$ and 
the ordinate is infrared luminosity in 10$^{10}$ L$_{\odot}$. 
Open circles are from \citet{gao04a}, and the filled square is our data
of NGC 4418.
The solid line is our adopted best fit line to the data by Gao \&
Solomon, assuming that infrared luminosity is proportional to HCN
luminosity (L$_{\rm IR}$ = 7.46 $\times$ L$_{\rm HCN}$ in the adopted
units). 
}
\end{figure}

\begin{figure}
\includegraphics[angle=-90,scale=0.45]{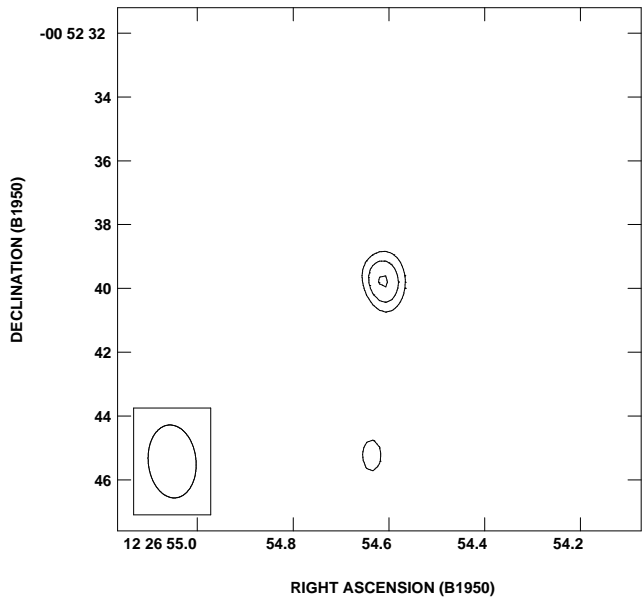}
\caption{
Contour map of continuum emission from NGC 4418 at $\lambda$ $\sim$ 3.4
mm ($\nu$ $\sim$ 88 GHz).
The beam size is 2$\farcs$3 $\times$ 1$\farcs$5 (position
angle is 5.8$^{\circ}$ east of the north).
Contours start at 5.6 mJy beam$^{-1}$ and increase in steps of 1.4 mJy
beam$^{-1}$.  
}
\end{figure}

\begin{figure}
\includegraphics[angle=-90,scale=.45]{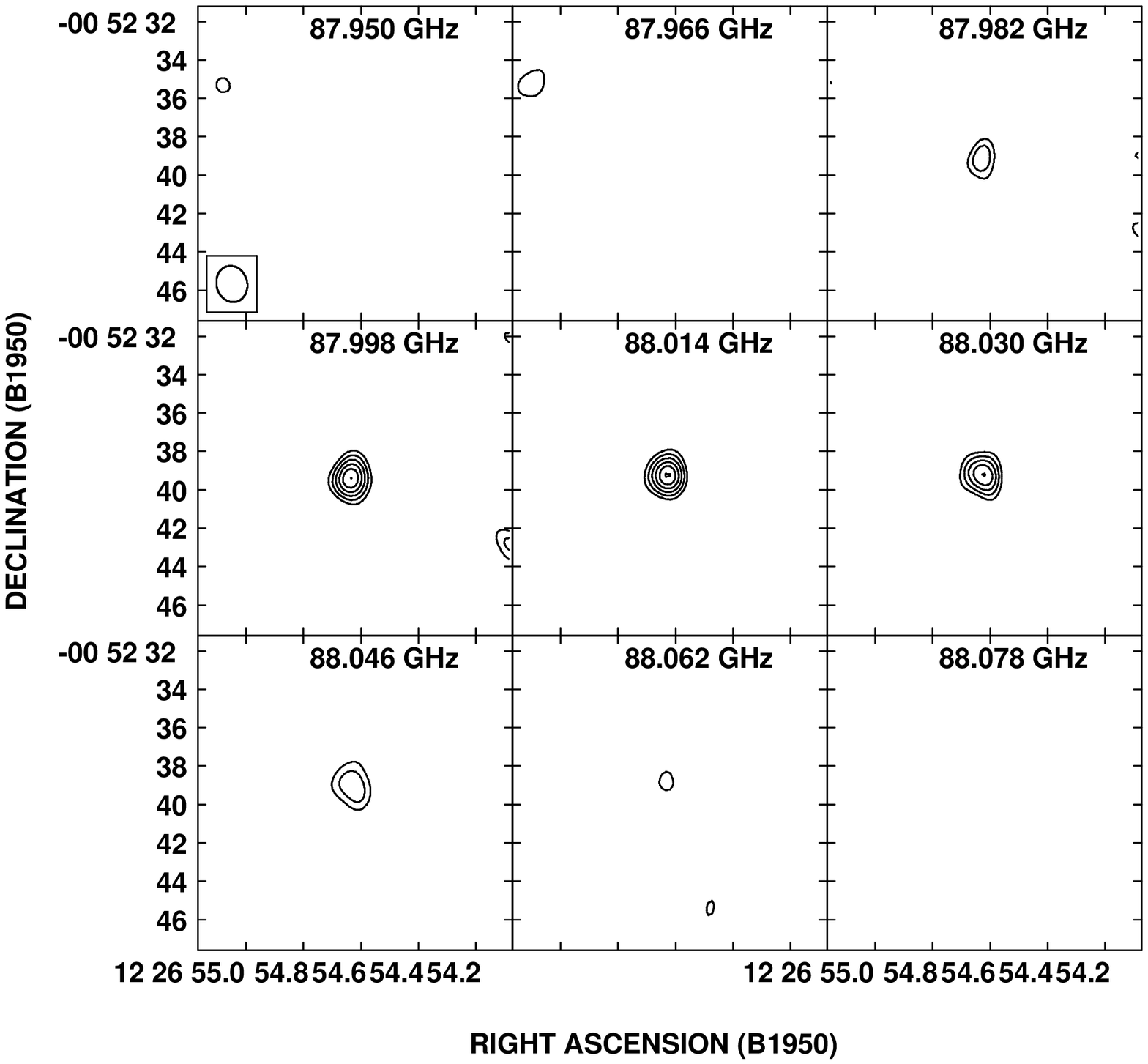}
\includegraphics[angle=-90,scale=.45]{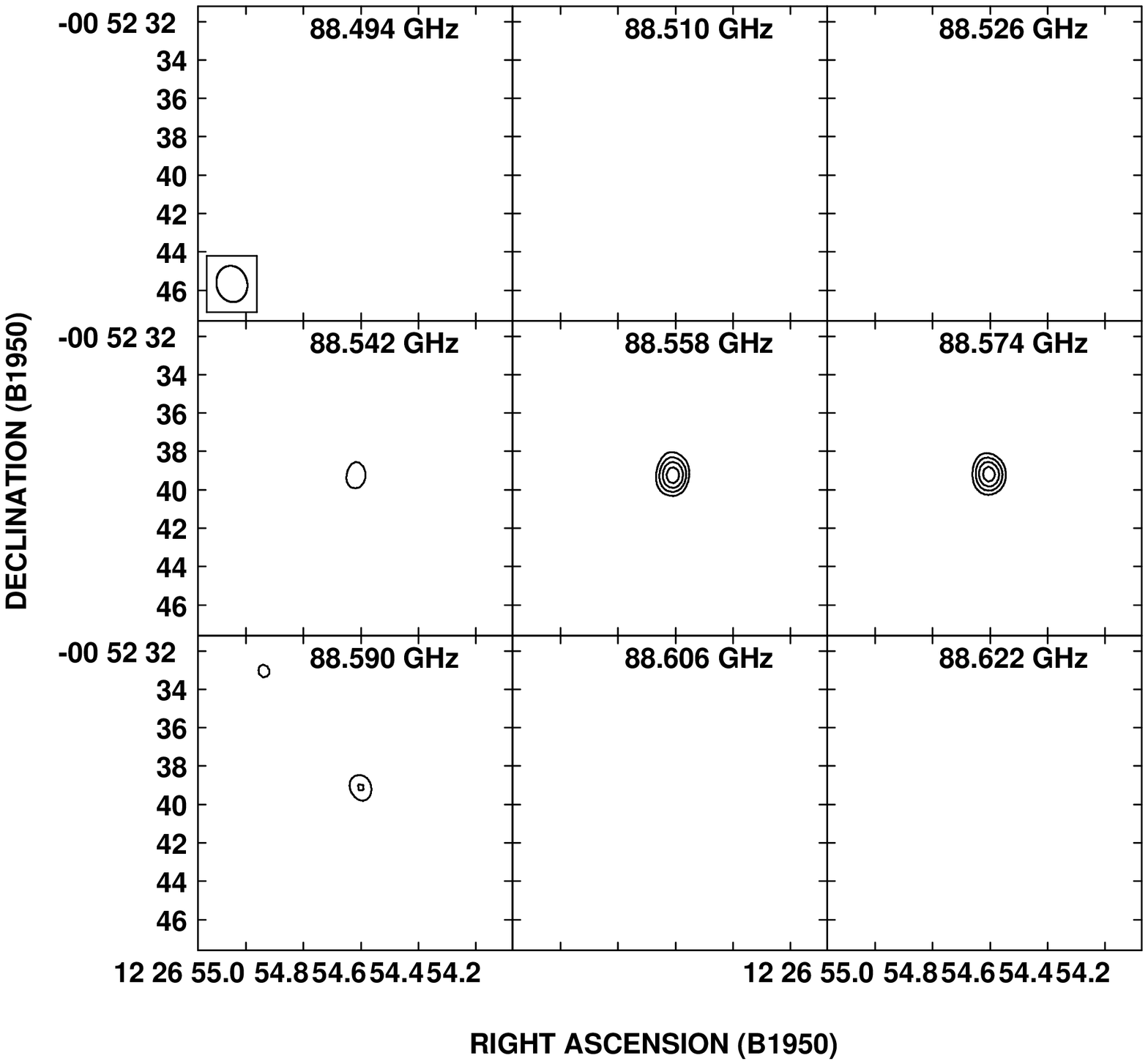}
\caption{
Channel maps of HCN (1--0) ({\it Left}) and HCO+ (1--0) ({\it Right})
emission lines obtained with the RAINBOW and NMA. 
The beam size at these lines is 1$\farcs$9 $\times$ 1$\farcs$6 (position
angle is 14.7$^{\circ}$ east of the north).
Contours start at 15 mJy beam$^{-1}$ and increase in steps of 5 
mJy beam$^{-1}$. 
}
\end{figure}

\begin{figure}
\includegraphics[angle=-90,scale=.6]{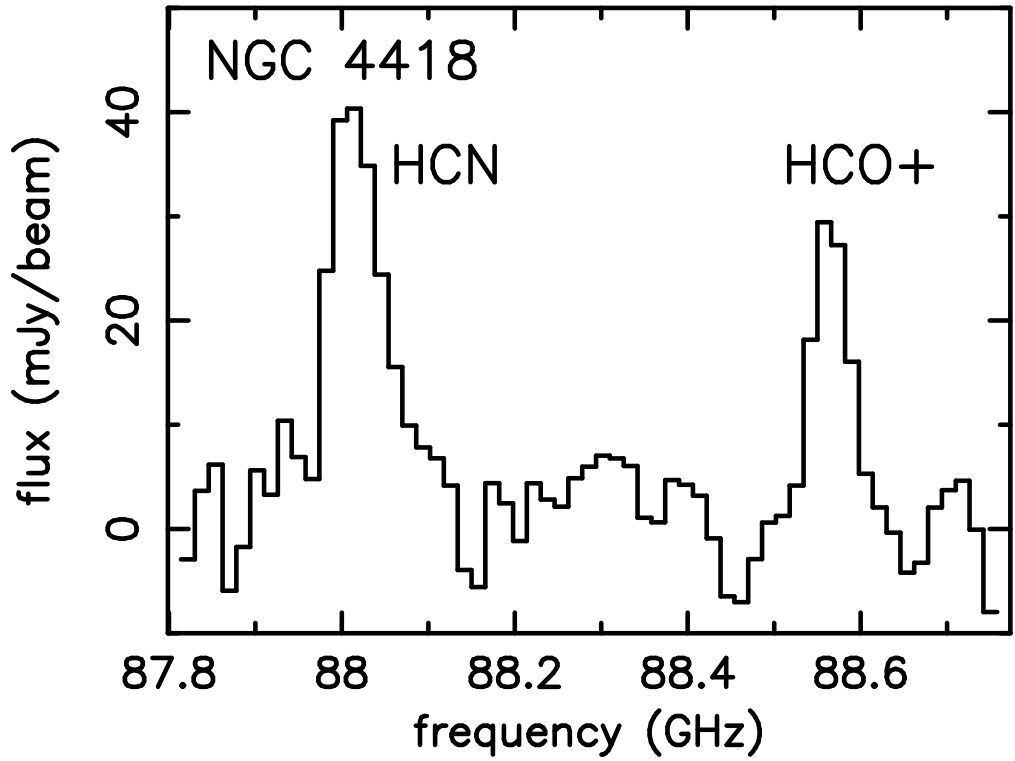} \\
\caption{
Nuclear spectrum of NGC 4418 around
the HCN (1--0) and HCO+ (1--0) lines, simultaneously obtained with the
RAINBOW and NMA.
The spectrum is continuum-subtracted, and the CLEAN procedure was
applied.
The abscissa is observed frequency in GHz, and the ordinate is flux in
mJy beam$^{-1}$. 
}
\end{figure}

\begin{figure}
\includegraphics[angle=-90,scale=.7]{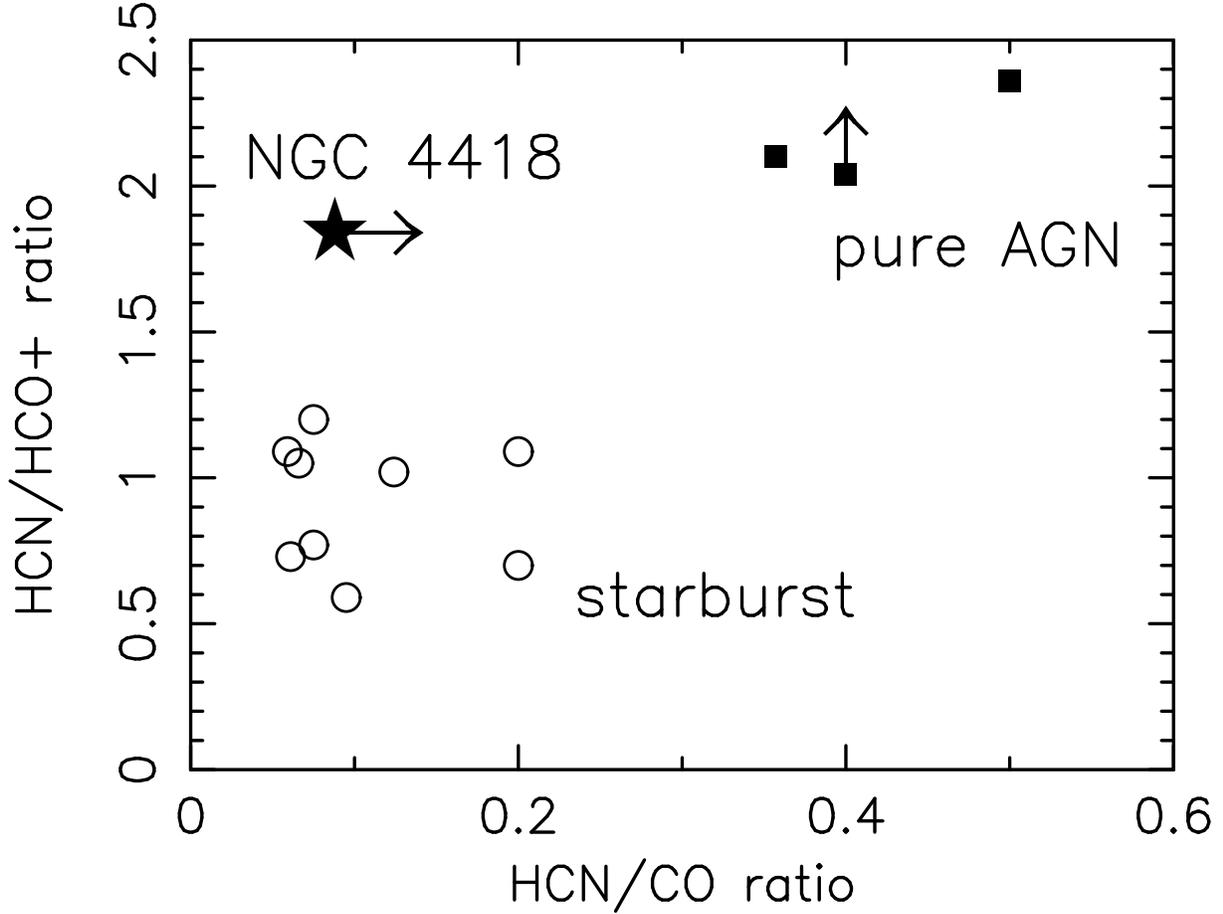}
\caption{
Plot of HCN (1--0) to CO (1--0) ($\lambda_{\rm rest}$ = 2.6 mm)
(abscissa) and HCN (1--0) to HCO+ (1--0) (ordinate)
luminosity ratios. 
The data other than NGC 4418 are taken from \citet{koh02}.
For NGC 4418, the HCN/HCO+ luminosity ratio is that directly measured 
toward the nucleus from our millimeter interferometric data.
For the HCN/CO ratio, we adopt the values based on single dish
telescopes' measurements of CO \citep{san91,yao03} and HCN (our NRO 45m
data), because no interferometric observations at the CO line have been
reported for NGC 4418.  
The single dish data reflect the global HCN/CO luminosity ratio from a
significant fraction of a host galaxy.
Since the molecular gas density is likely to be higher with  proximity 
to the nucleus, it is expected that the HCN/CO ratio at the nucleus 
should be higher than the adopted ratio.
Thus, the plotted HCN/CO should be taken as a lower limit for the actual
value at the core of NGC 4418. 
}
\end{figure}

\begin{figure}
\includegraphics[angle=-90,scale=.50]{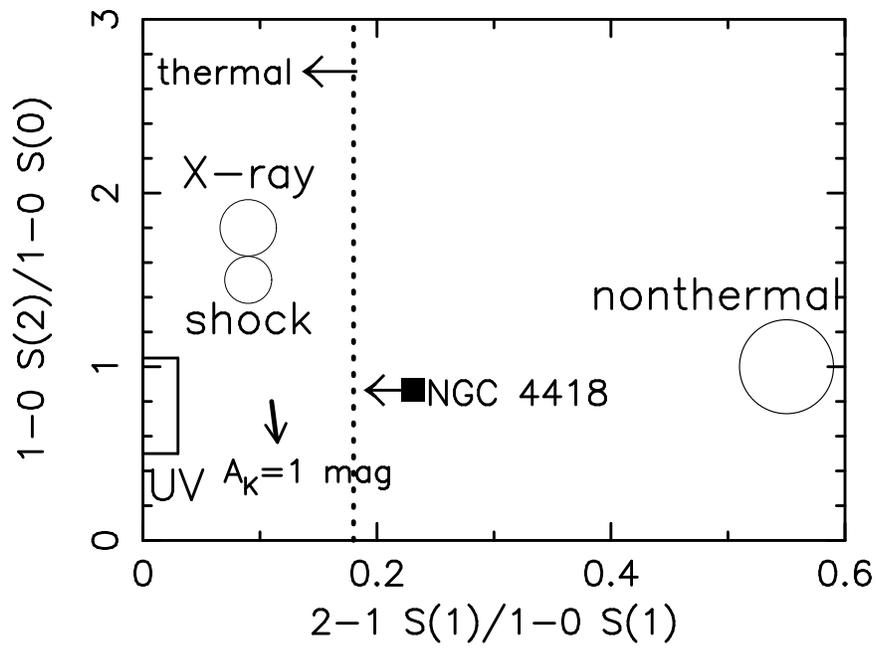}
\caption{
Luminosity ratios of H$_{2}$ emission lines in NGC 4418.
The abscissa is the 2--1 S(1) to 1--0 S(1) luminosity ratio, 
and the ordinate is the 1--0 S(2) to 1--0 S(0) luminosity ratio.
The typical ratios for non-thermal, shock, UV, and X-ray models are 
indicated \citep{mou94}. 
}
\end{figure}

\end{document}